\documentclass[pra]{revtex4}
\usepackage{amsmath}
\usepackage{graphicx}
\usepackage{amssymb}

 \newcommand{\ket}[1]{\left\vert #1\right\rangle}
\newcommand{\qed}{\hfill \ensuremath{\Box}\vspace{8pt}}
\newcommand\Id{\leavevmode\hbox{\small1\normalsize\kern-.33em1}}

\begin{document}
\title{Quantum Limits on Parameter Estimation}

\author{G. Goldstein$^{1}$, M. D. Lukin$^{1}$, P. Cappellaro$^{2}$}
\address{$^{1}$Department of Physics, Harvard University, Cambridge MA 02138 USA}
\address{$^{2}$Nuclear Science and Engineering Dept., Massachusetts Institute of Technology, Cambridge MA 02139 USA}

\begin{abstract}
We present a new proof of the quantum Cramer-Rao bound for precision
parameter estimation \cite{key-1,key-2,key-3} and extend it to a more general class of measurement procedures. 
We analyze a generalized framework for parameter estimation that covers most experimentally
accessible situations, where multiple rounds of measurements, auxiliary
systems or external control of the evolution are available. The proof
presented demonstrates the equivalence of these more general metrology
procedures to the simplest optimal strategy for which the bound is proven: a single measurement
of a two-level system interacting with a time-independent Hamiltonian. 
\end{abstract}
\maketitle

\section{Introduction}
High sensitivity parameter estimation is an active area of research
in quantum physics. There is growing effort both theoretically and
experimentally to use quantum properties of matter to improve the
precision with which a given parameter may be estimated. These ideas
have been used in several problems of practical interest; namely,
clock synchronization \cite{key-4,key-5,key-6,key-7,key-8}, reference
frame alignment \cite{key-9,key-10}, phase estimation \cite{key-11,key-12,key-13},
frequency measurements \cite{key-14,key-15,key-16,key-17}, position
measurements \cite{key-18,key-19} and magnetometry \cite{key-20,key-21}.
The simplest procedure for parameter estimation uses a probe which
is coupled to the external field ($b$) to be measured by a Hamiltonian
$b\, H$. The probe is prepared in a well-known initial state and
then interacts with the field for a time $\tau$ before the measurement
of a suitable observable $\mathcal{O}$. The process is then repeated
for a large number of times ($N$) to improve statistics.

Many different strategies have been proposed to improve the sensitivity
limit of the simple parameter estimation procedure~\cite{Giovannetti06}.
For example, the probe can be a composite system \cite{Wineland92}
or be augmented by ancillary systems used for multiple quantum non-demolition
(QND) measurements \cite{key-22}. The external field Hamiltonian
can be manipulated by additional field-independent and controllable
Hamiltonians to obtain an effective Hamiltonian $b\,\bar{H}$~\cite{key-14,key-15,key-16,key-17,key-20,key-21}. 
During the evolution time $\tau$ many positive operator valued
measurements (POVMs) can be performed and the results of the measurements
used in a feedback loop \cite{key-23,key-24}. 
The only constraint on the metrology procedure is 
that a single measurement time is limited to $\tau$. This assumption is physically
motivated as any measurement process suffers from decoherence that
limits the sensing time.

The quantum Cramer-Rao bound gives a  bound on the achievable sensitivity~\cite{key-1,key-2,key-3}.
 For any measurement scheme, if the largest and smallest eigenvalues
of $H$ are $\Lambda$ and $\lambda$, respectively, the optimum possible
sensitivity is bounded by: 
\begin{equation}
\delta b\geq\frac{1}{\tau\sqrt{N}\left(\Lambda-\lambda\right)},
\label{eq: Sensitivity Limit}
\end{equation}
where $N$ is the number of measurement runs and we set $\hbar=1$. This
is very similar to the Heisenberg limit for precision measurements
with entangled states where in the Heisenberg limit $\Lambda-\lambda$
is related to the number of entangled spins used for the quantum
measurement~\cite{key-26}. 

In this paper we present a new rigorous proof of this theorem. Our
approach is to reduce general parameter estimation problems often 
studied in the literature (involving e.g. larger systems, mixed states or POVMs) 
 to the case of a two-level system. Furthermore we extend the bound validity showing that  multiple
rounds of POVMs and feedback cannot improve this limit. 

We first prove in Section \ref{TLS} the sensitivity bound 
for a single POVM measurement on an isolated two-level system
in a pure state. The proof relies on the classical Fisher Information 
(reviewed in the Appendix), which provides
a lower bound on the uncertainty of parameter estimation via multiple
measurements in terms of the probabilities of various measurement outcomes. 
We then show in Section \ref{NLS} that for the purposes of precision
measurement a general $N$-level system prepared in a pure state
is equivalent to a two-level system. 
Specifically,  we will demonstrate an explicit reduction of the $N$-level system to one of its two dimensional subspaces; 
then extend these results to the case where a control Hamiltonian
is added to the field dependent
Hamiltonian ($H=b\, H+H_{0}$ ) by going to an appropriate interaction
picture. By using ``convexity'' properties of Fisher Information and
Cauchy-Schwartz inequalities, we  also prove in Section \ref{mixed}  the
bound for mixed states.

 In section \ref{feedback} we further prove that these results  are still valid when  
feedback during the measurement and classical communication
between different measurement rounds are available, 
situations where the Cramer-Rao bound has not been proved before.

Finally, in Section \ref{example}, we give an example of an experimentally accessible system where the
proven bound can be satisfied, before drawing our conclusions in Section \ref{conclusions}.

\section{Bound for a Single Two-Level System Prepared in a Pure State}
\label{TLS}
\textbf{\textsc{Lemma 1 --}} Consider parameter estimation using a
single two level system. Suppose that the system interacts with the 
Hamiltonian $b\cdot H$ (with largest and smallest
eigenvalues $\Lambda$ and $\lambda$ respectively) for a time $\tau$.
The system is initialized in the state $\left|\Psi_{in}\right\rangle $
and at the end of the sensing sequence an operator $\mathcal{O}$
is measured . The procedure is repeated $N$ times. Then, the
minimum uncertainty of $b$  is given by: 
\begin{equation}
{\displaystyle \inf_{\left|\Psi_{in}\right\rangle ,\mathcal{O}}\delta b=\frac{1}{\tau\sqrt{N}\left(\Lambda-\lambda\right)},}
\label{eq:TwoLevelSensitivity}
\end{equation}
 where  the infimum is taken over all initial states $\left|\Psi_{in}\right\rangle $ and observables
$\mathcal{O}$.

\textbf{Proof --} Given an operator $\mathcal{O}$, the precision
with which $b$ can be determined is given by: 
\begin{equation}
\delta b=\frac{\langle\Psi_{fin}|\Delta\mathcal{O}|\Psi_{fin}\rangle}{\sqrt{N}|\partial\langle\Psi_{fin}|\mathcal{O}|\Psi_{fin}\rangle/\partial b|}
\approx\frac{\left\langle \Psi_{in}\right|\Delta\mathcal{O}\left|\Psi_{in}\right\rangle }{\tau\sqrt{N}\left|\left\langle \Psi_{in}\right|\left[H,\mathcal{O}\right]\left|\Psi_{in}\right\rangle \right|},
\label{eq:Hamiltonian Sensitivity}
\end{equation}
 where $|\Psi_{fin}\rangle=e^{-ib\, Ht}\ket{\Psi_{in}}$ and the second line is obtained by first order perturbation theory.

First we show that the limit given by Eq. (\ref{eq:TwoLevelSensitivity})
above can be attained. Explicitly if we choose $\mathcal{O}=\left|\Lambda\right\rangle \left\langle \lambda\right|+\left|\lambda\right\rangle \left\langle \Lambda\right|$
and $\ket{\Psi_{in}} =\frac{1}{\sqrt{2}}\left(\left|\Lambda\right\rangle +i\left|\lambda\right\rangle \right)$,
we obtain $\frac{\left|\left\langle \Psi_{in}\right|\left[H,\mathcal{O}\right]\left|\Psi_{in}\right\rangle \right|}{\left\langle \Delta\mathcal{O}\right\rangle }=\Lambda-\lambda$.

To prove that this is the optimal bound we consider a general
initial state and measurement Hamiltonian. First we observe that Eq. (\ref{eq:TwoLevelSensitivity}) is
invariant under the substitutions $H\rightarrow\kappa H+\mu\Id$
and $\mathcal{O}\rightarrow\chi\mathcal{O}+\nu\Id$. As a result we can take $H=\overrightarrow{h}\cdot\overrightarrow{\sigma}$
and $\mathcal{O}=\overrightarrow{o}\cdot\overrightarrow{\sigma}$
(with $\left\Vert \overrightarrow{h}\right\Vert ,\left\Vert \overrightarrow{o}\right\Vert =1$). 
Because of rotational invariance of Eq. (\ref{eq:TwoLevelSensitivity}), without
loss of generality we can assume that $H=\frac{\sigma_{z}}{2}$
and $\mathcal{O}=\cos{\alpha}\,\sigma_{z}+\sin{\alpha}\,\sigma_{x}$,
with initial state $|\Psi_{in}\rangle=\cos{(\vartheta/2)}|0\rangle+e^{i\varphi/2}\sin{(\vartheta/2)}|1\rangle$.
Then $\langle\mathcal{O}\rangle=\cos{\alpha}\cos{\vartheta}+\sin{\alpha}\sin{\vartheta}\cos{\left(bt-\varphi/2\right)}$,
and since $\mathcal{O}^{2}=\Id$, the uncertainty in the external
field is given by \[
\delta b=\frac{\sqrt{1-\left[\cos{\alpha}\cos\theta+\sin\alpha\sin{\vartheta}\cos{\left(bt-\frac{\phi}{2}\right)}\right]^{2}}}{t\sin{\alpha}\sin{\vartheta}\sin\left(bt-\frac{\varphi}{2}\right)}.\]
 Taking the derivative with respect to $\alpha$ and $\vartheta$,
we find that the maximum is obtained $\forall\varphi$ for $\alpha=\pm\pi/2$
and $\vartheta=\pm\pi/2$ and it is equal to $\delta b=\frac{1}{\tau\sqrt{N}}$
(which matches Eq. (\ref{eq:TwoLevelSensitivity}) given that the
spread of eigenvalues of $\frac{\sigma_{z}}{2}$ is one).$\qed$

\textbf{\textsc{Lemma 2 --}} Consider parameter estimation using a
single two level system. Suppose the system interacts with the external
field via an effective Hamiltonian $b\, H$. The largest and smallest
eigenvalues of $H$ are $\Lambda$ and $\lambda$ respectively. The
system is initialized in a state $\left|\Psi_{in}\right\rangle $
and after a time $\tau$ a generalized measurement described by a
set of POVMs $\left\{ E_{\alpha}\right\} $ is performed. If this
procedure is repeated $N$ times, the minimum uncertainty of $b$
is: \begin{equation}
\delta b_{min}\equiv{\displaystyle \inf_{\left|\Psi_{in}\right\rangle ,\left\{ E_{\alpha}\right\} }\delta b=\frac{1}{\tau\sqrt{N}\left(\Lambda-\lambda\right)},}\label{eq:Two Level Sensitivity (POVM)}\end{equation}
 where the infimum is taken over all initial states $\left|\Psi_{in}\right\rangle $
and POVMs $\left\{ E_{\alpha}\right\} $ .

\textbf{Proof --} Let $\left\{ E_{1},\, E_{2},...E_{K}\right\} $
be any POVM, and $\left|\Psi_{in}\right\rangle $ any given initial
state. To first order in $b$, the probability of the measurement
outcome being $E_{\alpha}$ is given by $P\left(E_{\alpha}\right)=\left\langle \Psi_{fin}\right|E_{\alpha}\left|\Psi_{fin}\right\rangle \approx\left\langle \Psi_{in}\right|E_{\alpha}\left|\Psi_{in}\right\rangle +ib\tau\left\langle \Psi_{in}\right|\left[H,E_{\alpha}\right]\left|\Psi_{in}\right\rangle \equiv P_{0}\left(E_{\alpha}\right)+b\delta P\left(E_{\alpha}\right)$.
Then, by Lemma 3 in the Appendix (classical Fisher information), the
uncertainty in the external field is: 
\begin{equation}
\delta b_{min}^{2}=\left({N\sum\frac{\left(i\tau\left\langle \Psi_{in}\right|\left[H,E_{\alpha}\right]\left|\Psi_{in}\right\rangle \right)^{2}}{\left\langle \Psi_{in}\right|E_{\alpha}\left|\Psi_{in}\right\rangle }}\right)^{-1}
\label{eq:MinB(POVM)}
\end{equation}
Furthermore,  according to Sublemma 1 (see Eq. \ref{eq:Measurement Operator} in the Appendix) 
 the same sensitivity may be obtained by measuring 
the operator $\mathcal{O}\equiv\sum_\alpha\frac{E_{\alpha}}{P_{0}\left(E_{\alpha}\right)}$.
We have thus reduced the problem to the case where we measure  a
single operator and we may apply the results of Lemma 1 to obtain
the bound  (\ref{eq:Two Level Sensitivity (POVM)}). $\qed$

\section{Cramer-Rao Bound For Higher Dimensional Systems}
\label{NLS} 
We will now reduce parameter estimation with general
pure states to the two dimensional case studied in Lemma 1.

\textbf{\textsc{Proposition 1 --}} Consider parameter estimation with
an arbitrary probe in an $n$-dimensional Hilbert space. Suppose
that the system interacts with the external field via the 
Hamiltonian $b\, H$ (with largest and smallest eigenvalues $\Lambda$
and $\lambda$ respectively) for a time $\tau$. The system is initialized
in the state $\left|\Psi_{in}\right\rangle $ and at the end of the
sensing sequence a POVM measurement with operators $\left\{ E_{\alpha}\right\} $
is performed. The procedure is repeated $N$ times. Then the minimum
uncertainty $\delta b_{min}$  is given by Eq. (\ref{eq: Sensitivity Limit}).

\textbf{Proof --} We reformulate the $n$-dimensional problem in terms
of the two-dimensional case we just proved.

For any initial state $\left|\Psi_{in}\right\rangle $ we define $\left|\Omega_{in}\right\rangle \equiv H\left|\Psi_{in}\right\rangle $.
We can reduce the measurement procedure to a measurement on the subspace
$V_{S}$ spanned by $\left\{ \left|\Psi_{in}\right\rangle ,\,\left|\Omega_{in}\right\rangle \right\} $
since 
\begin{equation}
\begin{array}{ll}
\delta b_{min}^{2}&=\left(N\sum\frac{\left(ib\tau\left\{ \left\langle \Omega_{in}\right|E_{\alpha}\left|\Psi_{in}\right\rangle -\left\langle \Psi_{in}\right|E_{\alpha}\left|\Omega_{in}\right\rangle \right\} \right)^{2}}{\left\langle \Psi_{in}\right|E_{\alpha}\left|\Psi_{in}\right\rangle }\right)^{-1}\\
&=\left({N\sum\frac{\left(ib\tau\left\{ \left\langle \Omega_{in}\right|\Pi E_{\alpha}\Pi\left|\Psi_{in}\right\rangle -\left\langle \Psi_{in}\right|\Pi E_{\alpha}\Pi\left|\Omega_{in}\right\rangle \right\} \right)^{2}}{\left\langle \Psi_{in}\right|\Pi E_{\alpha}\Pi\left|\Psi_{in}\right\rangle }}\right)^{-1},\end{array}\label{eq:precisionPOVMManyDim}\end{equation}
 where $\Pi$ is the projector onto the space spanned by
$\left|\Omega_{in}\right\rangle $ and $\left|\Psi_{in}\right\rangle $.
When restricted to the two dimensional subspace spanned by $\{\left|\Psi_{in}\right\rangle ,\ \left|\Omega_{in}\right\rangle \}$
the set of operators $\left\{ \Pi E_{\alpha}\Pi\right\} $ still forms
a POVM, since all the operators are positive definite $\left\langle \Psi\right|\Pi E_{\alpha}\Pi\left|\Psi\right\rangle \geq0$
and $\sum\Pi E_{\alpha}\Pi=\Id_{2}$ (where $\Id_{2}$ is
the identity on the subspace). Furthermore the spread of the Hamiltonian's eigenvalues
 ($\Lambda-\lambda$) cannot increase when restricted to a
smaller subspace. We can thus apply the results given in Lemma 2 to
conclude that optimum sensitivity is given by Eq. (\ref{eq: Sensitivity Limit}).$\qed$

\textbf{\textsc{Corollary 1 --}} \textit{Bound for additional control
Hamiltonians.} \\
 Consider parameter estimation using an arbitrary probe in a pure
state $\left|\Psi_{in}\right\rangle $. Suppose the system evolves
for a time $\tau$ with the Hamiltonian $b\, H+H_{0}(t)$,
before a POVM $\left\{ E_{\alpha}\right\} $ is performed. If the
sensing sequence is repeated $N$ times, the minimum uncertainty of
$b$ over all states is given by Eq. (\ref{eq:Two Level Sensitivity (POVM)}).

\textbf{Proof --} To prove the bound we write the evolution of the
system in the interaction picture defined by the Hamiltonian $H_{0}$.
The evolution is then given by $U=U_{0}^{\dag}U_{H}^{int}U_{0}$,
where to leading order in $b\tau$ we can write the propagator $U_{H}^{int}\approx e^{-i\bar{H}^{int}\tau}$
in terms of the average Hamiltonian $\bar{H}^{int}=\frac{1}{\tau}\int H^{int}(t)dt$, 
with $H^{int}(t)=U_{0}^{\dag}(t)HU_{0}(t)$. 
By applying Proposition 2 to the initial state $U_{0}\left|\Psi_{in}\right\rangle $ and defining
$\Lambda^{int},\ \lambda^{int}$ the largest and smallest eigenvalues
of $\bar{H}^{int}$, the optimum sensitivity is given by: $\delta b_{min}=\frac{1}{\sqrt{N}\tau(\Lambda^{int}-\lambda^{int})}$.

To prove the bound we now only need to show that $|\Lambda^{int}-\ \lambda^{int}|\leq\Lambda-\lambda$ . 
To this goal we first rephrase this condition in terms of the norm of $\bar{H}^{int}$.
The well-known  equivalence~\cite{key-25} between the operator (or
spectral) norm $\|~\|_{2}$ and the Frobenius norm $\|~\|_{F}$ for
Hermitian operators,  $\|H\|_{F}=\|H\|_{2}$, implies that 
$\max\left\{ \left|\Lambda\right|,\,\left|\lambda\right|\right\}=\displaystyle \sup_{|\Psi\rangle}|\langle\Psi|H|\Psi\rangle|$.

Without loss of generality we may set the smallest eigenvalue of
$H$ to zero, so it is sufficient to show that the magnitude of the
largest eigenvalue of $\bar{H}^{int}$ is less then that of $H$ and
all eigenvalues stay positive. 
Since $\forall\,\ket{\psi}$  we have
\begin{equation}
|\langle\psi|\bar{H}^{int}|\psi\rangle|\leq\frac{1}{\tau}\int_{0}^{\tau}dt|\left\langle \psi\right|U_{0}^{\dag}(t)HU_{0}(t)\left|\psi\right\rangle |\leq\left\Vert H\right\Vert, 
\label{eq:Operator Norms}\end{equation}
the largest magnitude eigenvalue of $\bar{H}^{int}$
is less than the eigenvalues spread of $H$. 
Also, since $\langle\psi|\bar{H}^{int}|\psi\rangle=\frac{1}{\tau}\int_{0}^{\tau}dt\left\langle \psi\right|U_{0}^{\dag}(t)HU_{0}(t)\left|\psi\right\rangle \geq0$, $\forall\,\ket{\psi}$,   all the eigenvalues of $\bar{H}^{int}$ are positive, proving that the spread of eigenvalues of $\bar{H}^{int}$
is less then that of $H$.

We thus proved that the sensitivity cannot be improved beyond the limit
given by Eq. (\ref{eq: Sensitivity Limit}) by adding a 
time-dependent control Hamiltonian. $\qed$

\section{Mixed states}\label{mixed} 

\textbf{\textsc{Proposition 2 --}} \textit{Bound for mixed states.} \\
 Consider the same scenario as in Corollary I, but now the system
is initialized in the mixed state $\rho_{in}$. The minimum uncertainty
of $b$ over all mixed states is still given by the Cramer-Rao bound, Eq. (\ref{eq: Sensitivity Limit}).

\textbf{Proof --} Following Corollary 1, we can always eliminate $H_{0}$
in the interaction picture by replacing $\rho$ with $\rho^{int}=U_{0}^{\dag}\rho U_{0}$
and $H$ with $\overline{H_{int}}$. Thus without loss of generality
we can assume $H_{0}=0$. In this case from the initial state $\rho\left(0\right)=\sum P_{i}\left|\Psi_{i}\right\rangle \left\langle \Psi_{i}\right|$
we have $\rho\left(\tau\right)=\sum P_{i}\left|\Psi_{i}+\delta\Psi_{i}\right\rangle \left\langle \Psi_{i}+\delta\Psi_{i}\right|$.
To leading order, the probability of an outcome $E_{\alpha}$ is then
$P\left(E_{\alpha}\right)\cong\sum P_{i}\left\{ \left\langle \Psi_{i}\right|E_{\alpha}\left|\Psi_{i}\right\rangle +\left\langle \delta\Psi_{i}\right|E_{\alpha}\left|\Psi_{i}\right\rangle +\left\langle \Psi_{i}\right|E_{\alpha}\left|\delta\Psi_{i}\right\rangle \right\} $.
Using Lemma 2 in the appendix (classical Fisher Information) we can
express the sensitivity as a function of the measurement probabilities:
\begin{equation}
\delta b_{min}^{2}=\left[N\sum_{\alpha}\frac{\left(\sum_i P_{i}\left\{ \left\langle \delta\Psi_{i}\right|E_{\alpha}\left|\Psi_{i}\right\rangle +\left\langle \Psi_{i}\right|E_{\alpha}\left|\delta\Psi_{i}\right\rangle \right\} \right)^{2}}{\sum_i P_{i}\left\langle \Psi_{i}\right|E_{\alpha}\left|\Psi_{i}\right\rangle }\right]^{-1}
\label{eq:Sensitivity}
\end{equation}
Applying the Cauchy-Schwartz inequality to $\left\{\sum_i P_{i}\left[\left\langle \delta\Psi_{i}\right|E_{\alpha}\left|\Psi_{i}\right\rangle +\left\langle \Psi_{i}\right|E_{\alpha}\left|\delta\Psi_{i}\right\rangle \right] \right\}^{2}$, we have
\[  \left\{ \sum_{i}\left[\left(\frac{\sqrt{P_{i}}\left(\left\langle \delta\Psi_{i}\right|E_{\alpha}\left|\Psi_{i}\right\rangle +\left\langle \Psi_{i}\right|E_{\alpha}\left|\delta\Psi_{i}\right\rangle \right)}{\sqrt{\left\langle \Psi_{i}\right|E_{\alpha}\left|\Psi_{i}\right\rangle }}\right)\sqrt{P_{i}\left\langle \Psi_{i}\right|E_{\alpha}\left|\Psi_{i}\right\rangle }\right] \right\}^2 \leq \]
\[\left(\sum_{i}\frac{P_{i}\left(\left\langle \delta\Psi_{i}\right|E_{\alpha}\left|\Psi_{i}\right\rangle +\left\langle \Psi_{i}\right|E_{\alpha}\left|\delta\Psi_{i}\right\rangle \right)^{2}}{\left\langle \Psi_{i}\right|E_{\alpha}\left|\Psi_{i}\right\rangle }\right) \left(\sum_{i}P_{i}\left\langle \Psi_{i}\right|E_{\alpha}\left|\Psi_{i}\right\rangle \right)\]
Then, following Proposition 1 and changing the order of summation we obtain 
\begin{equation}
\delta b_{min}^{2}\leq\frac{1}{N\sum_{\alpha}\sum_{i}P_{i}\frac{\left(\left\langle \delta\Psi_{i}\right|E_{\alpha}\left|\Psi_{i}\right\rangle +\left\langle \Psi_{i}\right|E_{\alpha}\left|\delta\Psi_{i}\right\rangle \right)^{2}}{\left\langle \Psi_{i}\right|E_{\alpha}\left|\Psi_{i}\right\rangle }}\leq\frac{1}{N\sum_{i}P_{i}\tau^{2}\left(\Lambda-\lambda\right)^{2}}=\frac{1}{N\tau^{2}\left(\Lambda-\lambda\right)^{2}},
\label{eq:MultilineOptimization}
\end{equation}
 showing  that a mixture of pure states is less efficient then a single pure
state.  Incidentally,
this also demonstrates the ``convexity'' of Fisher information~\cite{key-1,key-2}.$\qed$

Note that
since $\sum P_{i}\frac{\left(\left\langle \delta\Psi_{i}\right|E_{\alpha}\left|\Psi_{i}\right\rangle +\left\langle \Psi_{i}\right|E_{\alpha}\left|\delta\Psi_{i}\right\rangle \right)^{2}}{\left\langle \Psi_{i}\right|E_{\alpha}\left|\Psi_{i}\right\rangle }\leq sup_{i}\frac{\left(\left\langle \delta\Psi_{i}\right|E_{\alpha}\left|\Psi_{i}\right\rangle +\left\langle \Psi_{i}\right|E_{\alpha}\left|\delta\Psi_{i}\right\rangle \right)^{2}}{\left\langle \Psi_{i}\right|E_{\alpha}\left|\Psi_{i}\right\rangle }$
given any density matrix we can always find one of its pure state
components that provides a better initial state for quantum metrology.

We now assume that an ancillary system (or a partially controllable
environment) is available. We show that even with these added resources,
the sensitivity bound does not improve.\\
 \textbf{\textsc{Corollary 2 --}} \textit{Bound for mixed states
coupled to an ancillary system.} \\
 Suppose that the system interacts for a time $\tau$ with the
external field and an ancillary system via the Hamiltonian $b\, H+H_{0}+H_{a}$, 
where $H_{a}$ does not depend on $b$, but includes the interaction between sensor and ancillas.
The system is initialized in the state $\rho_{in}$ and at the end
of the sensing sequence a POVM measurement $\left\{ E_{\alpha}\right\} $
is performed on the system. If the procedure is repeated $N$
times then the minimum uncertainty of $b$ is given
by Eq. (\ref{eq: Sensitivity Limit}).

\textbf{Proof --} Consider the system composed by the ancillary system 
and the probe. The extension of the POVMs $\left\{ E_{\alpha}\right\} $
to this larger system $\left\{ E_{\alpha}\otimes\Id\right\} $
via the identity on the ancillas is still a POVM. We thus reduced
the problem to proposition 2. $\qed$

We would like to note that if the ancilla
Hamiltonian  $H_{a}$ were $b$-dependent the bound could be violated.  
In that case, the probe plus ancillas can be considered as a single system with a new sensing Hamiltonian
$H'=H+H_a$ that can have a larger spread of eigenvalues than $H$. 
An example where the effect of the external field on the ancillas 
is used to enhance sensitivity is given in Section \ref{example}.

\section{Feedback}\label{feedback} 
We will now include the possibility of multiple 
rounds of POVM measurements, first with feedback only during each round (Proposition 3) and then allowing classical communication
between measurement rounds (Proposition 4). 
These propositions extend the known results~\cite{key-1,key-2,key-3}, for which we gave  new proofs in the previous sections, 
to more general and inclusive metrology procedures, proving that the bound in Eq. (\ref{eq: Sensitivity Limit}) is still optimal.

\begin{figure}[htb]
\centering
\includegraphics[scale=0.6]{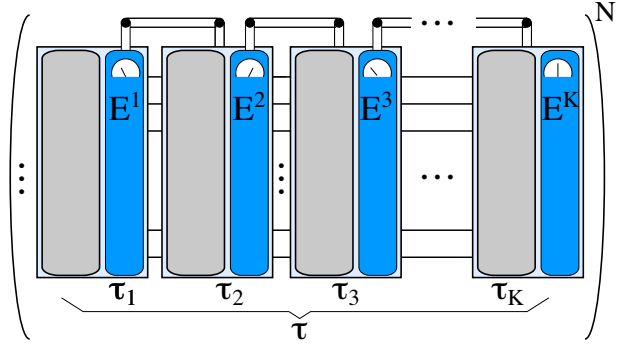} 
\caption{\label{cap:FeedbackScheme} Multiple measurement and feedback scheme.
The probe system (pictured as multiple qubits for simplicity) interacts
with the external field and the control Hamiltonian during $K$ intervals each of length $\tau_{i}$
for a total time $\tau$ (gray rectangles). After each period a POVM measurement ($\left\{ E_{\alpha}^{i}\right\} $)
is performed on the system. Feedback is applied between each one of the $K$ steps 
based on the previous measurement outcome. The same scheme is then repeated  $N$ times to improve statistics.}
\end{figure}

\textbf{\textsc{Proposition 3 --}} \textit{Bound for mixed states with feedback.}\\
 Suppose that the system is initialized in state $\rho_{in}$ and evolves under the Hamiltonian
$b\, H+H_{0}(t)$. The evolution is interrupted by the measurement of  sets of POVMs $\left\{ E_{i}^{\alpha}\right\} $. 
The control Hamiltonian $H_0(t)$  and the POVMs are chosen using feedback based on the previous measurement results. 
The overall measurement procedure lasts a
time $\tau$ and is repeated $N$ times to improve statistics (see Fig. \ref{cap:FeedbackScheme}). Then, the
minimum uncertainty of $b$  is given by Eq. (\ref{eq: Sensitivity Limit}): 
 $\delta b_{min}\geq\frac{1}{\tau\sqrt{N}\left(\Lambda-\lambda\right)}$.

\textbf{Proof --} By inserting  identity
operators as POVMs  at appropriate times,  we may assume that every
experiment run consists of $K$ measurements at times $\left\{ \tau_{1},\,\tau_{2},....\tau_{K-1},\,\tau\right\} $
with POVMs given by $\left\{ E_{\alpha}^{1}\right\} ,\,\left\{ E_{\alpha}^{2}\right\} ,....\left\{ E_{\alpha}^{K}\right\} $
respectively. Following the strategy used to prove Proposition 1,
we would like to eliminate the explicit feedback loop and external
Hamiltonian. 
For this purpose let the POVM be $E_{\beta}^{m}=\left(M_{\beta}^{m}\right)^{\dagger}M_{\beta}^{m}$
and the unitary evolution conditioned by feedback on the outcome $\left\{ E_{\alpha_{1}}^{1},\, E_{\alpha_{2}}^{2},...E_{\alpha_{m-1}}^{m-1}\right\} $
be $U_{\alpha_{1},\alpha_{2},...,\alpha_{m-1}}$. By replacing $M_{\beta}^{m}$ with  $U_{\alpha_{1},\alpha_{2},...,\alpha_{m-1}}M_{\beta}^{m}$
we can reproduce the feedback by applying a different set of POVM measurements. 
Also, we can set $H_0=0$ by going to the interaction picture with respect to $H_{0}$ and 
replacing $\rho_{in}$ with $e^{iH\tau_{1}}\rho_{in}e^{-iH\tau_{1}}$ and 
$M_{\alpha}^{L}$ with  $e^{-iH_{0}\left(\tau_{L+1}-\tau_{L}\right)}M_{\alpha}^{L}$
(also $bH$  becomes time-dependent in the interaction picture). 
Overall any procedure involving feedback
and a control Hamiltonian is equivalent to a different POVM and a
time dependent $bH\left(t\right)$. We thus want to prove that this
cannot give a better bound than the optimal POVM strategy.

Now we wish to calculate various uncertainties (see the Appendix) in terms
of probabilities of various measurement outcomes. For zero external
field the probability of the outcome $\left\{ E_{\alpha_{1}}^{1},\, E_{\alpha_{2}}^{2},...E_{\alpha_{K}}^{K}\right\} $
is given by:
\begin{equation}
P_{0}\left\{ E_{\alpha_{1}}^{1},\, E_{\alpha_{2}}^{2},...E_{\alpha_{K}}^{K}\right\} =Tr\left\{ \rho_{in}E_{\alpha_{K}..\alpha_{1}}\right\}, 
\label{eq:Probability Trace}
\end{equation}
where $E_{\alpha_{K}..\alpha_{L}}\equiv\left(M_{\alpha_{K}}...M_{\alpha_{L}}\right)^{\dagger}M_{\alpha_{K}}...M_{\alpha_{L}}$.
For non-zero external field,  to leading
order in $b$ the change in the probability of a given outcome $P\left\{ E_{\alpha_{1}}^{1},\, E_{\alpha_{2}}^{2},...E_{\alpha_{K}}^{K}\right\} $
is:
\begin{equation}
\delta P\left\{ E_{\alpha_{1}}^{1},...,E_{\alpha_{K}}^{K}\right\} \cong\sum_{L=0}^{K-1}ib\left(\tau_{L}-\tau_{L-1}\right)\cdot Tr\left\{ \left[A_{\alpha_{L}...\alpha_{1}}\rho_{in}\left(A_{\alpha_{L}..\alpha_{1}}\right)^{\dagger},\overline{H_{L}}\right]E_{\alpha_{K}..\alpha_{L+1}}\right\}, 
\label{eq:ChangeProbability}
\end{equation}
where $A_{\alpha_{L}..\alpha_{1}}\equiv M_{\alpha_{L}}...M_{\alpha_{1}}$
and $\overline{H_{L}}=\frac{\int_{\tau_{L}}^{\tau_{L+1}}H\left(t\right)}{\tau_{L+1}-\tau_{L}}$.
Using the classical Fisher Information formulas given in the appendix
we may write that:
\[
\begin{array}{l}
\left(\delta b_{min}^{2}\right)^{-1}=N\sum_{\alpha_{1},\alpha_{2},..\alpha_{K}}\frac{\left(\delta P\left\{ E_{\alpha_{1}}^{1},\, E_{\alpha_{2}}^{2},...E_{\alpha_{K}}^{K}\right\} \right)^{2}}
{b^2\,P_{0}\left\{ E_{\alpha_{1}}^{1},\, E_{\alpha_{2}}^{2},...E_{\alpha_{K}}^{K}\right\} }=\\
=N \displaystyle{\sum_{L,M=0}^{K-1}}\sum_{\alpha_{1},\alpha_{2},..\alpha_{K}}\frac{\left(\tau_{L+1}-\tau_{L}\right)\cdot\left(\tau_{M+1}-\tau_{M}\right)}{Tr\left\{ \rho_{in}E_{\alpha_{K}....\alpha_{1}}\right\} }Tr\left\{ \left[A_{\alpha_{L}..\alpha_{1}}\rho_{in}\left(A_{\alpha_{L}...\alpha_{1}}\right)^{\dagger},\overline{H_{L}}\right]E_{\alpha_{K}..\alpha_{L+1}}\right\}\\
\qquad\qquad\qquad\qquad\cdot Tr\left\{ \left[A_{\alpha_{L}..\alpha_{1}}\rho_{in}\left(A_{\alpha_{L}...\alpha_{1}}\right)^{\dagger},\overline{H_{L}}\right]E_{\alpha_{K}...\alpha_{M+1}}\right\}
\end{array}
\]
 In the second step we have changed the order of summation. Applying
the Cauchy-Schwartz inequality to the sum $\sum_{\alpha_{1}..\alpha_{K}}$
(to separate contributions corresponding to different POVM measurements)
we obtain 
\[
\begin{array}{l}
\delta b_{min}^{-2}\leq N\displaystyle{\sum_{L,M=0}^{K-1}}\left\{\left(\tau_{L+1}-\tau_{L}\right)\left(\tau_{M+1}-\tau_{M}\right)\left[\sum_{\alpha_{1}..\alpha_{K}}\frac{Tr\left\{ \left[A_{\alpha_{L}...\alpha_{1}}\rho_{in}A_{\alpha_{L}..\alpha_{1}}^{\dagger},\overline{H_{L}}\right]E_{\alpha_{K}..\alpha_{L+1}}\right\} ^{2}}{Tr\left\{ \rho_{in}E_{\alpha_{K}...\alpha_{1}}\right\} }\right]^{1/2}\right.\\
\left.\qquad\qquad\qquad\qquad\:\left[\sum_{\alpha_{1}...\alpha_{K}}\frac{Tr\left\{ \left[A_{\alpha_{L}..\alpha_{1}}\rho_{in}A_{\alpha_{L}..\alpha_{1}}^{\dagger},\overline{H_{L}}\right]E_{\alpha_{K}...\alpha_{M+1}}\right\} ^{2}}{Tr\left\{ \rho_{in}E_{\alpha_{K}....\alpha_{1}}\right\} }\right]^{1/2}\right\}
\end{array}\label{eq:CauchySchwatz}
\]
 Since the operator $A_{\alpha_{L}..\alpha_{1}}\rho_{in}A_{\alpha_{L}..\alpha_{1}}^{\dagger}\equiv s\rho_{\alpha_{L}..\alpha_{1}}$
is positive definite, up to a scaling factor $s$ it represents a density operator.
Also we note that 
\[
Tr\left\{ \rho_{in}E_{\alpha_{K}....\alpha_{1}}\right\} =Tr\left\{ A_{\alpha_{L}..\alpha_{1}}\rho_{in}A_{\alpha_{L}..\alpha_{1}}^{\dagger}E_{\alpha_{K}..\alpha_{L+1}}\right\} 
\]
 and that $\left\{ E_{\alpha_{K}....\alpha_{L+1}}\right\} $ is a
POVM. As a result, we can apply Proposition 3 to the normalized   $A_{\alpha_{L}..\alpha_{1}}\rho_{in}A_{\alpha_{L}..\alpha_{1}}^{\dagger}$, obtaining 
\begin{equation}
\begin{array}{l}
\sum_{\alpha_{1}..\alpha_{K}}\frac{\left(Tr\left\{ \left[A_{\alpha_{L}..\alpha_{1}}\rho_{in}A_{\alpha_{L}..\alpha_{1}}^{\dagger},\overline{H_{L}}\right]E_{\alpha_{K}..\alpha_{L+1}}\right\} \right)^{2}}{Tr\left\{ \rho_{in}E_{\alpha_{K}....\alpha_{1}}\right\} }\\
\leq\sum_{\alpha_{1}..\alpha_{L}}\left(\Lambda-\lambda\right)Tr\left\{ A_{\alpha_{L}..\alpha_{1}}\rho_{in}A_{\alpha_{L}..\alpha_{1}}^{\dagger}\right\} =\Lambda-\lambda\end{array}\label{eq:SpecificInequalities}\end{equation}
 The first inequality derives from Proposition 2 and the fact that the
spread of eigenvalues of $\overline{H}$ is less then $\Lambda-\lambda$.
The last equality is obtained by noting that $\sum_{\alpha_{1}..\alpha_{L}}A_{\alpha_{L}..\alpha_{1}}^{\dagger}A_{\alpha_{L}..\alpha_{1}}=\Id$.
Finally we obtain the bound
\begin{equation}
\delta b_{min}^{-2}\leq N{\displaystyle \sum_{L=0}^{K-1}\sum_{M=0}^{K-1}}\left(\tau_{L+1}-\tau_{L}\right)\left(\tau_{M+1}-\tau_{M}\right)\left(\Lambda-\lambda\right)^{2}=N\tau^{2}\left(\Lambda-\lambda\right)^{2}
\label{eq:FinalUncertainty}
\end{equation}
 We therefore conclude that multiple POVM rounds and feedback cannot
improve the sensitivity beyond the limit given by Eq. (\ref{eq: Sensitivity Limit}).$\qed$

Note that by choosing a set of POVMs $\{E_{\alpha_{1}},...,E_{\alpha_{L}}\}$ that maximizes
the sum $\sum_{\alpha_{K}..\alpha_{L+1}}\frac{\left(Tr\left\{ \left[\rho_{\alpha_{L}..\alpha_{1}},\overline{H_{L}}\right]E_{\alpha_{K}..\alpha_{L+1}}\right\} \right)^{2}}{Tr\left\{ \rho_{\alpha_{L}..\alpha_{1}}E_{\alpha_{K}....\alpha_{L+1}}\right\} }$
we can find a single step in the multiple POVM sequence that is at least
as efficient as the entire  feedback sequence. 

\begin{figure}[hbt]
\includegraphics[scale=0.4]{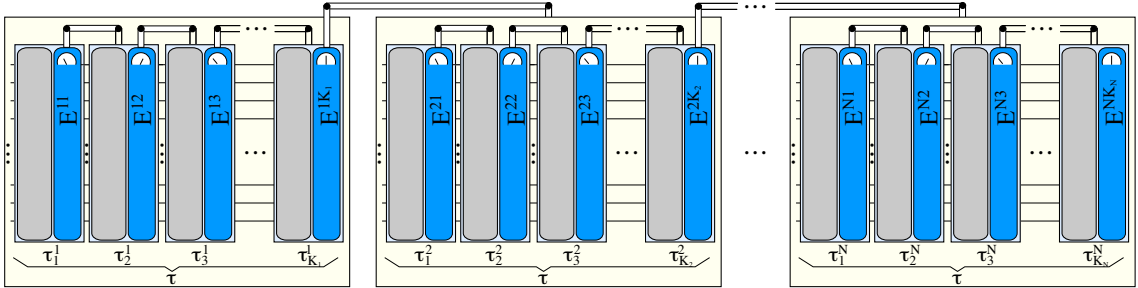} 
\caption{\label{cap:Feedback&Control} Multiple round measurement and feedback
scheme (with classical communication between rounds). The probe system
(pictured as multiple qubits for simplicity) undergoes $N$ measurement
rounds each lasting a total time $\tau$. The evolution in  each round $i$ is 
subdivided into $K_{i}$ intervals each of
length $\tau_{j}^{i}$. During each interval, the system  
interacts with the external field and a control Hamiltonian (gray rectangle) that depends on feedback from the previous interval and the previous round. 
After each time interval, a POVM measurement $\left\{ E_{\alpha}^{i,j}\right\} $ (chosen according to the feedback scheme)
is performed on the system (blue rectangle). The result of the measurement is used to control the next time interval or the next measurement round. }
\end{figure}

\textbf{\textsc{Proposition 4 --}} \textit{Bound for mixed states
with feedback and multi-round measurements.}\\
 Suppose the system is initialized in the state $\rho_{in}$ and interacts with the Hamiltonian
$b\, H+H_{0}$. 
During the evolution a set of POVMs $\left\{ E_{\mu}^{1,i}\right\} $
are measured . Here $i$ stands for the POVM measurement number,
$\mu$ is the outcome and $1$ identifies the first round of measurements.
Feedback based on the measurement outcomes determines the control Hamiltonian and the choice of 
 POVMs. The overall measurement procedure lasts a time $\tau$. The
next round of measurement uses a potentially different initial state,  a different set of POVMs $\left\{ E_{\nu}^{2,j}\right\} $ and 
a different feedback scheme. Furthermore the second measurement procedure
may depend on the results of the first measurement and  also lasts a time $\tau$. 
A  total of $N$
rounds of measurements are carried out, so  that the total measurement time is given by $N\tau$ (see Fig. \ref{cap:Feedback&Control}).
The minimum uncertainty of $b$  obtained by this scheme is given by Eq. (\ref{eq: Sensitivity Limit}), 
$\delta b_{min}\geq\frac{1}{\tau\sqrt{N}\left(\Lambda-\lambda\right)}$.

\textbf{Proof} -- By Corollary 4 (see Appendix) we know that $\delta b_{min}^{2}\geq\left[{\sum_{1\rightarrow N}\frac{\left(\delta P\left(O_{\alpha}^{1\rightarrow N}\right)\right)^{2}}{P_{0}\left(O_{\alpha}^{1\rightarrow N}\right)}}\right]^{-1}$,
where the sum is over all possible outcomes $O_{\alpha}^{1\rightarrow N}$
of the $N$ rounds of POVM measurements. We wish to prove by induction
on $N$ that $\left(\delta b_{min}^{2}\right)^{-1}\leq N\tau^{2}\left(\Lambda-\lambda\right)^{2}$.
The case $N=1$ is given by Proposition 3. 
If we assume that $\sum_{1\rightarrow N-1}\frac{\left(\delta P\left(O_{\alpha}^{1\rightarrow N-1}\right)\right)^{2}}{P_{0}\left(O_{\alpha}^{1\rightarrow N-1}\right)}\leq\left(N-1\right)\tau^{2}\left(\Lambda-\lambda\right)^{2}$, we obtain  the bound:
\begin{equation}
\begin{array}{l}
\sum_{1\rightarrow N}\frac{\left(\delta P\left(O_{\alpha}^{1\rightarrow N}\right)\right)^{2}}{P_{0}\left(O_{\alpha}^{1\rightarrow N}\right)}=\sum_{1\rightarrow N-1}\sum_{N}\left(\frac{\left(\delta P\left(O_{\beta}^{1\rightarrow N-1}\right)\cdot P_{0}\left(O_{\gamma}^{N}\right)+\delta P\left(O_{\beta}^{1\rightarrow N-1}\right)\cdot P_{0}\left(O_{\gamma}^{N}\right)\right)^{2}}{P_{0}\left(O_{\alpha}^{1\rightarrow N-1}\right)\cdot P_{0}\left(O_{\gamma}^{N}\right)}\right)\\
=\sum_{1\rightarrow N-1}\sum_{N}\left\{ \frac{\left(\delta P\left(O_{\beta}^{1\rightarrow N-1}\right)\right)^{2}\cdot P_{0}\left(O_{\gamma}^{N}\right)}{P_{0}\left(O_{\beta}^{1\rightarrow N-1}\right)}+\frac{\left(\delta P\left(O_{\alpha}^{N}\right)\right)^{2}\cdot P_{0}\left(O_{\beta}^{1\rightarrow N-1}\right)}{P_{0}\left(O_{\alpha}^{N}\right)}\right\} \leq N\tau^{2}\left(\Lambda-\lambda\right)^{2}\end{array}\label{eq:MultiroundInequalities}\end{equation}
 Here the outcome $O_\alpha^{1\rightarrow N}$ is given by the outcomes $O_{\beta}^{1\rightarrow N-1}$ in the
first $N-1$ rounds of measurement and $O_\gamma$ in the last round. 
The first equality holds because the probability of the last measurement outcome is independent of the previous
measurements. 
The second equality derives from  $\sum\delta P\left(O_{\alpha}^{N}\right)=0$. Finally, by noting that 
$\sum P_{0}\left(O_{\alpha}^{N}\right)=\sum P_{0}\left(O_{\beta}^{1\rightarrow N-1}\right)=1$, we can obtain by induction the last inequality.$\qed$

This result indicates that classical communication between different
measurement rounds cannot improve sensitivity beyond the limit given
in Eq. (\ref{eq: Sensitivity Limit}). Specifically, the ``independence''
of the uncertainties between steps of the multi-round  strategy (demonstrated
in Eq.~\ref{eq:MultiroundInequalities}) indicates that the sensitivity
obtained by choosing one of the measurement rounds is at least as
high as that of the overall procedure.

\section{Example: Sensitivity Improvement with Auxiliary Qubits}\label{example}
We now present an illustration of the bounds derived in this paper, in particular
the effects of an ancillary system and an external control field. 
In many experimental situations \cite{key-20,key-21}
a probe consists of a quantum sensor (for simplicity a two-level system)
and a spin environment. The external field, which we wish to measure,
is coupled to both the sensor and the environment.  
The sensitivity of the probe can then be enhanced by using the environment
spins as ancillas to enhance the response of the system to the external
field. 

We assume that the sensor spin (which can be prepared in a well defined initial
state, coherently manipulated and read out) is coupled to a bath of
``dark'' spins, which can be polarized and collectively controlled
but cannot be directly detected. 
The system is described by the Hamiltonians:
\begin{equation}
\begin{array}{l}
H=H_{meas}+H_{int}\\
H_{int}=\left|1\right\rangle \left\langle 1\right|\lambda{\displaystyle \sum I_{x}^{i}},\: H_{meas}=  b\left(\left|1\right\rangle \left\langle 1\right|+{\displaystyle \sum I_{z}^{i}}\right),
\end{array}
\label{eq:Main Hamiltonian}
\end{equation}
 where $\lambda$ is the coupling between the sensor and environment
spins. Here $\left|0\right\rangle ,\,\left|1\right\rangle $ refer
to the sensor spin while $\left|\uparrow\right\rangle ,\,\left|\downarrow\right\rangle ,\, I_{z}^{i}$
describe the dark spins. 
We shall consider the case where $H_{int}$ can
be turned on and off at will and is much larger in magnitude then
any other interaction in the system. As the spread of eigenvalues
of $H_{meas}$ is equal to $1+K$ (where $K$ is the total number
of ancillary spins) in principle it should be possible to attain Heisenberg
limited metrology (with sensitivity scaling $\sim\frac{1}{K }$)
using this Hamiltonian. This is very similar to metrology using GHZ
states or systems with multi-body coupling to the parameter~\cite{key-27}. 
\begin{figure}[htbp]
\centering \includegraphics[scale=1]{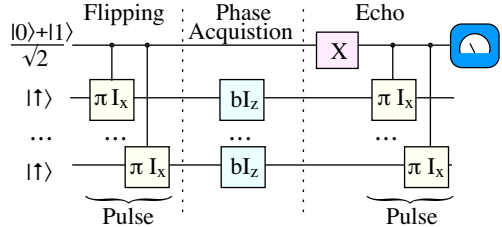} 
\caption{\label{fig:CentralSpin}A quantum circuit used to enhance parameter
estimation sensitivity. CNOT gates make the state of the dark spins
dependent on the state of the sensor spin. The dark spins pick up
different phases dependent on the state of the sensor spin and the
echo followed by more CNOT gates maps this phase onto the sensor
spin which is then read out. Here $\tau_{pulse}=\frac{\pi}{2\lambda}$,
$X$ stands for a $\pi$ pulse on the sensor qubit flipping $\left|0\right\rangle \leftrightarrow\left|1\right\rangle $.}
\end{figure}

To illustrate this method we consider the idealized case when the
coupling between the sensor spin and the dark spins are in our control
and the dark spins are initialized in a pure state: $\left|\uparrow...\uparrow\right\rangle $.
Consider the circuit shown in Fig. \ref{fig:CentralSpin}. First,
the sensor spin is prepared in an equal superposition of the two
internal states $\left|0\right\rangle +\left|1\right\rangle $ (dropping
normalization). Then $H_{int}$ (CNOT gates on the dark spins) is
used to produce the state: \begin{equation}
\left|0\right\rangle \left|\uparrow...\uparrow\right\rangle +\left|1\right\rangle \left|\downarrow...\downarrow\right\rangle ,\label{eq: Initial state}\end{equation}
 This state is then used to sense the magnetic field. The action of
the external field leads to the states $\left|0\right\rangle \left|\uparrow...\uparrow\right\rangle$ and $\left|1\right\rangle \left|\downarrow...\downarrow\right\rangle $
acquiring different phases $ \tau\frac{K+2}{2}$ and $- \tau\frac{K}{2}$, respectively.
After the interaction with the magnetic field, the sensor spin is flipped
and another control operation with $H_{int}$ is applied. This leads
to the following final state for the total spin system:
\begin{equation}
\left(e^{-i \tau b\left(1+\frac{K}{2}\right)}\left|1\right\rangle +e^{\frac{i \tau bK}{2}}\left|0\right\rangle \right)\left|\downarrow...\downarrow\right\rangle 
\label{eq:Final Wavefunction}
\end{equation}
 Note that this is a product state of the
sensor spin and the dark spin states. If we then measure  the operator
$\mathcal{O\equiv}i\left(\left|0\right\rangle \left\langle 1\right|-\left|1\right\rangle \left\langle 0\right|\right)$
(say $N$ times to improve statistics) we would get a minimum uncertainty
$b_{min}\doteq\sqrt{\frac{1}{N}}\frac{1}{ \left(1+K\right)\tau}$
(or Heisenberg limited metrology). This effect may be understood by
noting that the circuit shown in Fig. 2 effectively converts the measurement
Hamiltonian $H_{meas}\equiv b\left(\left|1\right\rangle \left\langle 1\right|+{\displaystyle \sum I_{z}^{i}}\right)$ to a new interaction $\widetilde{H}_{meas} \equiv b\left|1\right\rangle \left\langle 1\right|\left(\mathbb{I}+{\displaystyle \sum I_{z}^{i}}\right)$.
This new Hamiltonian  is much more convenient, since it is possible to prepare the optimal initial state, 
 $\frac{1}{\sqrt{2}}\left(\left|0\right\rangle +\left|1\right\rangle \right)\left|\downarrow...\downarrow\right\rangle $ (which 
is a an equal superposition of the two eigenstates with largest and smallest eigenvalues) and measure the  optimal operator
for this state and Hamiltonian, $i\left(\left|0\right\rangle \left\langle 1\right|-\left|1\right\rangle \left\langle 0\right|\right)$ (see Corollary 3 in the Appendix).

\section{Conclusions}\label{conclusions}

In this work we have presented a new proof of the Cramer-Rao bound and extended the bound 
to more general metrology frameworks, encompassing e.g. feedback. Key to our proof was the
realization that more complex metrology schemes cannot improve on the ideal parameter estimation 
performed via a two-level systems with the optimal initial state and observable.
Using only Cauchy-Schwartz inequalities and the Fisher Information, we proved that 
 the sensitivity cannot increase, even when adding external control, ancillary systems, using mixed states and 
 POVM measurements as well as multiple rounds of measurements with feedback.

Specifically, we systematically increased the complexity of the metrology procedure, by introducing 
one by one  additional features often considered in the literature and in experiments. 
The new metrology scheme obtained at each step is composed of many sub-procedures that have been 
considered in the previous scheme; we were  thus always able  to identify a sub-procedure that 
provided as high a sensitivity as the more complex metrology scheme. 
By backward induction, it is then possible to explicitly construct a two-level system, initialized
in a pure state, with no control Hamiltonians and a single operator
measurement that is a sub-step of the more complex measurement procedure
but is as efficient --or more-- than the whole measurement process.
Although this ideal simple system is not always experimentally accessible, 
and thus more complex strategies need to be adopted in practice, 
we showed that when constrained by a given sensing time, 
none of these strategies can surpass the fundamental Cramer-Rao limit.

\appendix
\section{ General Sensitivity Formulas}

\label{appendix}

For sake of completeness we present  a known result  ---the classical Fisher Information~\cite{key-1,key-2,key-3}---
that has been  used extensively in the main text of the paper. 
We show that maximum likelihood estimates
saturate the Fisher information bound in the limit of an infinite number of measurements
and we demonstrate the bound for finite number of measurements.

\textbf{\textsc{Lemma 3 --}} \textit{Generic bounds for parameter
estimation}.\\
 Consider a generic system coupled to some external field. The
system interacts with the field and potentially other control Hamiltonians.
It is possible that multiple sensing sequences are carried out on
the system; that is several sets of POVMs $\left\{ E_{\mu}^{1}\right\} ...\left\{ E_{\nu}^{m}\right\} $
are measured. The process is repeated $N$ times to improve statistics.
Suppose that $P\left(O_{\alpha}\right)\cong P_{0}\left(O_{\alpha}\right)+b\delta P\left(O_{\alpha}\right)$
(that is non-linear terms are negligible) where $P_{0}\left(O_{\alpha}\right)$
is the probability of measuring outcome $\alpha$ (which may be the
result of several POVMs) for zero external field. Then in the limit
$N\rightarrow\infty$ the minimum uncertainty for measuring the external
field is given by the classical Fisher Information:
\begin{equation}
\delta b_{min}^{2}=\left[{N\sum\frac{\left(\delta P\left(O_{\alpha}\right)\right)^{2}}{P_{0}\left(E_{\alpha}\right)}}\right]^{-1}
\label{eq:General uncertainty formula}
\end{equation}

Furthermore if only one POVM measurement $\left\{ E_{\alpha}\right\} $
is made, this sensitivity can also be obtained by measuring  the
operator:
\begin{equation}
\mathcal{O}\equiv\sum\frac{E_{\alpha}}{P_{0}\left(E_{\alpha}\right)}
\label{eq:Measurement Operator}
\end{equation}

\textbf{Proof --} We will begin by calculating the probabilities of
various outcomes of the measurements. After $N$ samplings the probability
of observing $F_{\alpha}$ frequency of each of the possible outcomes
$O_{\alpha}$ is given by:\begin{equation}
P\left(F_{1},...F_{K}\right)=\frac{N!}{\left(F_{1}\cdot N\right)!\cdot\cdot\cdot\left(F_{K}\cdot N\right)!}\Pi_{\alpha=1}^{K}\left(P\left(O_{\alpha}\right)\right)^{NF_{\alpha}}\cong Exp\left[-\frac{N}{2}\Sigma\frac{\left(P\left(O_{\alpha}\right)-F_{\alpha}\right)^{2}}{P\left(O_{\alpha}\right)}\right]\label{eq:ProbabilityOutcomes}\end{equation}
 Now we wish to isolate the part of Eq. (\ref{eq:ProbabilityOutcomes})
that depends on the external field $b$. Substituting $\Delta_{\alpha}\equiv F_{\alpha}-P_{0}\left(E_{\alpha}\right)$
we may write that: \begin{equation}
\begin{array}{l}
P\left(\Delta_{1},...\Delta_{K}\right)\cong Exp\left[-\frac{N}{2}\Sigma\frac{\left(P\left(O_{\alpha}\right)-F_{\alpha}\right)^{2}}{P\left(O_{\alpha}\right)}\right]=Exp\left[-\frac{N}{2}\Sigma\frac{\left(\frac{\Delta_{\alpha}}{\delta P\left(O_{\alpha}\right)}-b\right)^{2}\left(\delta P\left(O_{\alpha}\right)\right)^{2}}{P_{0}\left(O_{\alpha}\right)}\right]=\\
Exp\left[-\frac{N}{2}\sum\frac{\left(\delta P\left(E_{\alpha}\right)\right)^{2}}{P_{0}\left(E_{\alpha}\right)}\left(b-\frac{\sum\frac{\delta P\left(O_{\alpha}\right)}{P_{0}\left(O_{\alpha}\right)}\cdot\Delta_{\alpha}}{\sum\frac{\left(\delta P\left(O_{\alpha}\right)\right)^{2}}{P_{0}\left(O_{\alpha}\right)}}\right)^{2}\right]\cdot Exp\left[\sum\frac{\Delta_{\alpha}^{2}}{P_{0}\left(E_{\alpha}\right)}-\frac{\left[\sum\frac{\delta P\left(O_{\alpha}\right)}{P_{0}\left(O_{\alpha}\right)}\cdot\Delta_{\alpha}\right]^{2}}{\sum\frac{\left(\delta P\left(O_{\alpha}\right)\right)^{2}}{P_{0}\left(O_{\alpha}\right)}}\right]\end{array}\label{eq:GaussianUncertainty}\end{equation}

Qualitatively the second exponential after the last equality has no
$b$ dependence so cannot provide any further information about the
external field. This statement is made more quantitative in the following
sublemma.

\textbf{\textsc{Sublemma}} \textbf{1 --} The best possible estimate
for $b$ is given by $b_{opt}\cong\frac{\sum\frac{\delta P\left(E_{\alpha}\right)\Delta_{\alpha}}{P_{0}\left(E_{\alpha}\right)}}{\sum\frac{\left(\delta P\left(E_{\alpha}\right)\right)^{2}}{P_{0}\left(E_{\alpha}\right)}}$
(which is the maximum likelihood estimate) with uncertainty $\delta b_{opt}\cong\frac{1}{\sqrt{N\sum\frac{\left(\delta P\left(E_{\alpha}\right)\right)^{2}}{P_{0}\left(E_{\alpha}\right)}}}$.
For a single POVM $\left\{ E_{\alpha}\right\} $ this estimate can
be obtained by measuring the expectation of the operator $b=\left\langle \widetilde{\mathcal{O}}\right\rangle \equiv\frac{1}{\sum\frac{\left(\delta P\left(E_{\alpha}\right)\right)^{2}}{P_{0}\left(E_{\alpha}\right)}}\left\langle \sum\left(\frac{E_{\alpha}}{P_{0}\left(E_{\alpha}\right)}-1\right)\right\rangle $.
Note that $\widetilde{\mathcal{O}}$ is the same as $\mathcal{O}$
in Eq. (\ref{eq:Measurement Operator}) up to a constant and rescaling.

\textbf{Proof --} We note that the expectation of the operator is
indeed $b$ see Eq. (\ref{eq:GaussianUncertainty}). For fixed $b$
the expectation value of the operator $\widetilde{\mathcal{O}}$ comes
from a Gaussian distribution centered at $b$ of width $\frac{1}{\sqrt{N\sum\frac{\left(\delta P\left(O\alpha\right)\right)^{2}}{P_{0}\left(O_{\alpha}\right)}}}$.
As such we see that $\delta b_{opt}\leq\delta b_{\mathcal{O}}=\frac{1}{\sqrt{N\sum\frac{\left(\delta P\left(O_{\alpha}\right)\right)^{2}}{P_{0}\left(O_{\alpha}\right)}}}$.
We would like to show that this is indeed optimal. Let $S$ be any
statistic for $b$, that is a map of the frequency set onto $b$:
$S:\left(F_{1},...F_{K}\right)\rightarrow b$. The uncertainty for
this statistic is given by:\begin{equation}
\begin{array}{l}
\delta S^{2}=\lim_{L\rightarrow\infty}\frac{1}{2L}\intop_{-L}^{L}db\iiint dF_{1}..dF_{K}P\left(F_{1}..F_{K}\mid b\right)\cdot\left(b-S\left(F_{1},...F_{K}\right)\right)^{2}=\\
\lim_{L\rightarrow\infty}\frac{1}{2L}\iiint dF_{1}..dF_{K}\intop_{-L}^{L}db\cdot G\left(F_{1}..F_{K}\right)\cdot\left(b-S\left(F_{1},...F_{K}\right)\right)^{2}\cdot Exp\left[-\frac{\Lambda}{2}\left(b-\Delta\right)^{2}\right]=\\
\lim_{L\rightarrow\infty}\frac{1}{2L}\iiint dF_{1}..dF_{K}\intop_{-L}^{L}db\cdot G\left(F_{1}..F_{K}\right)\cdot\left(\frac{1}{\Lambda}+\left(b-S\left(F_{1},...F_{K}\right)\right)^{2}\right)\geq\frac{1}{\Lambda}\end{array}\label{eq:UncertaintyCalculation}\end{equation}

Here $\Delta={\sum\frac{\delta P\left(O\alpha\right)\Delta_{\alpha}}{P_{0}\left(O_{\alpha}\right)}}/{\sum\frac{\left(\delta P\left(O_{\alpha}\right)\right)^{2}}{P_{0}\left(O_{\alpha}\right)}}$,
$G\left(F_{1}..F_{K}\right)=Exp\left[\sum\frac{\Delta_{\alpha}^{2}}{P_{0}\left(O_{\alpha}\right)}-\frac{\left[\sum\frac{\delta P\left(O_{\alpha}\right)}{P_{0}\left(O\alpha\right)}\cdot\Delta_{\alpha}\right]^{2}}{\sum\frac{\left(\delta P\left(O_{\alpha}\right)\right)^{2}}{P_{0}\left(O_{\alpha}\right)}}\right]$
and $\Lambda=N\sum\frac{\left(\delta P\left(O_{\alpha}\right)\right)^{2}}{P_{0}\left(O_{\alpha}\right)}$
(see Eq. (\ref{eq:GaussianUncertainty})). In the first step we have
changed the order of integration, in the second we have used well
know properties of Gaussian integrals and for the third note that
$\left(b-S\left(F_{1},...F_{K}\right)\right)^{2},\ G\left(F_{1}..F_{K}\right)\geq0$.
In particular for one POVM measurement any statistic no more efficient
then measuring  $\widetilde{\mathcal{O}}$ or equivalently $\mathcal{O}$.
$\qed$

We thus proved the uncertainty bound in Eq. (\ref{eq:General uncertainty formula}).$\qed$

\textbf{Corollary 3 --}\textit{Optimal observable}.\\ 
Consider parameter estimation using the hypothesis in Corollary 1. The Cramer-Rao bound,  Eq. (\ref{eq: Sensitivity Limit}), cannot be violated by measuring an operator instead of a POVM  but it can be saturated by the measurement of a single observable  $\mathcal{O}$, for appropriate initial states. 

\textbf{Proof --}  First, measuring an operator cannot be more efficient than measuring a POVM, as for any operator a POVM made of its eigenvalues is completely equivalent.
Second, given an operator $\mathcal{O}$, the precision with
which $b$ can be determined is given by: 
\begin{equation}
\delta b=\frac{\Delta\mathcal{O}}{\sqrt{N}|\partial Tr\left\{ \rho\mathcal{O}\right\} /\partial b|}\approx\frac{\Delta\mathcal{O}}{\tau\sqrt{N}\left|Tr\left\{ \rho\left[H,\mathcal{O}\right]\right\} \right|},
\label{eq:Hamiltonian Sensitivity}
\end{equation}
 where the second line is obtained by first order perturbation theory
and $\Delta\mathcal{O}\equiv\sqrt{Tr\left\{ \rho\mathcal{O}^{2}\right\} -\left(Tr\left\{ \rho\mathcal{O}\right\} \right)^{2}}$.
Explicitly if we choose $\mathcal{O}=i\left|\Lambda\right\rangle \left\langle \lambda\right|-i\left|\lambda\right\rangle \left\langle \Lambda\right|$
and $\left|\Psi\right\rangle =\frac{1}{\sqrt{2}}\left(\left|\Lambda\right\rangle +\left|\lambda\right\rangle \right)$,
($\rho=\left|\Psi\right\rangle \left\langle \Psi\right|$) we obtain
$\frac{\left|\left\langle \Psi\right|\left[H,\mathcal{O}\right]\left|\Psi\right\rangle \right|}{\left\langle \Delta\mathcal{O}\right\rangle }=\Lambda-\lambda$.$\qed$

\textbf{Corollary 4 --} \textit{Generic bounds for parameter estimation
with finite number of trials. Consider a generic system coupled to
some external field.}\\
 The system interacts with the field and potentially other control
Hamiltonians. Possibly multiple sensing sequences are carried out
on the system that is several sets of POVMs $\left\{ E_{\alpha}^{1}\right\} ...\left\{ E_{\alpha}^{k}\right\} $
are measured. The process is repeated $K$ times to improve statistics.
Suppose that $P\left(O_{\alpha}\right)\cong P_{0}\left(O_{\alpha}\right)+b\delta P\left(O_{\alpha}\right)$
where $P_{0}\left(O_{\alpha}\right)$ is the probability of measuring
outcome $\alpha$ (which may be the result of several POVMs) for zero
external field. Then for any $K$ measurements the minimum uncertainty
for measuring the external field is given by:\begin{equation}
\delta b_{min}^{2}\geq\frac{1}{K\sum\frac{\left(\delta P\left(O_{\alpha}\right)\right)^{2}}{P_{0}\left(O_{\alpha}\right)}}\label{eq:General uncertainty formula finite}\end{equation}

Here $K$ is a finite number of repetitions of the experiment used
to improve statistics. In particular if the measurement is carried
out only once $\delta b_{min}^{2}\geq\frac{1}{\sum\frac{\left(\delta P\left(O_{\alpha}\right)\right)^{2}}{P_{0}\left(O_{\alpha}\right)}}$.

\textbf{Proof --} Consider any statistic ($S$) used to determine
$b$ using $K$ measurements, let it have uncertainty $\Delta_{K}$.
Now consider repeating this experiment $N\rightarrow\infty$ times
(for a total of $N\cdot K$ measurements). By Lemma 2 we know that
the optimum measurement produces uncertainty $\delta b_{opt}=\frac{1}{\sqrt{NK\sum\frac{\left(\delta P\left(O_{\alpha}\right)\right)^{2}}{P_{0}\left(O_{\alpha}\right)}}}$.
On the other hand taking the average of $N$ copies of statistic $S$
leads to uncertainty $\delta b_{opt}\leq\delta b_{NK}=\frac{1}{\sqrt{N}\Delta_{K}}$.
From this we see that $\Delta_{K}\geq\frac{1}{\sqrt{K\sum\frac{\left(\delta P\left(O_{\alpha}\right)\right)^{2}}{P_{0}\left(O_{\alpha}\right)}}}$
and Eq. (\ref{eq:General uncertainty formula finite}) follows. $\qed$

\textbf{Acknowledgments --} This work was supported by NSF and the Packard
Foundation. P.C. was in part supported by an ITAMP fellowship.



\begin{thebibliography}{99}
\bibitem{key-1} Braunstein S L and Caves C M 1994 \textit{Phys. Rev.
Lett.} \textbf{72} 3439 

\bibitem{key-2} Braunstein S L, Caves C M and Milburn G J 1996 \textit{Ann.
Phys. (N.Y.)} \textbf{247} 135 

\bibitem{key-3}Cramer H 1946 \textit{Mathematical Methods of Statistics}
(Princeton: Princeton University Press)

\bibitem{key-4} Jozsa R, Abrams D S, Dowling J P and Williams C P
2000 \textit{Phys. Rev. Lett.} \textbf{85} 2010 

\bibitem{key-5} Chuang I L 2000 \textit{Phys. Rev. Lett.} \textbf{85}
2006

\bibitem{Wineland92} Wineland D J, Bollinger J J, Itano W M, Moore
F L and Heinzen D J 1992 \textit{Phys. Rev. A} \textbf{46} R6797 

\bibitem{Giovannetti06} Giovannetti V, Lloyd S and Maccone L 2006
\textit{Phys. Rev. Lett}. \textbf{96} 010401

\bibitem{key-6} Revzen M and Mann A 2003 \textit{Phys. Lett. A} \textbf{312}
11 

\bibitem{key-7} de Burgh M and Bartlett S D 2005 \textit{Phys. Rev.
A} \textbf{72} 042301

\bibitem{key-8} Boixo S, Caves C M, Datta A and Shaji A 2006\textit{
Laser Phys.} \textbf{16} 1525 

\bibitem{key-9} Bagan E, Baig M and Tapia R M 2001 \textit{Phys.
Rev. Lett.} \textbf{87} 257903

\bibitem{key-10} Chiribella G, D'Ariano G M, Perinotti P and Sacchi
M F 2004 \textit{Phys. Rev. Lett.} \textbf{93} 180503

\bibitem{key-11} Gerry C C and Campos R A 2003 \textit{Phys. Rev.
A} \textbf{68} 025602 

\bibitem{key-12} Dunningham J A and Burnett K 2004 \textit{Phys.
Rev. A} \textbf{70} 033601

\bibitem{key-13} Wang H and Kobayashi T 2005 \textit{Phys. Rev. A}
\textbf{71} 021802(R)

\bibitem{key-14} Rosenband T, Hume D B, Schmidt P O, Chou C W, Brusch
A, Lorini L, Oskay W H, Drullinger R E, Fortier T M, Stalnaker J E
\textit{et al}. 2008 \textit{Science} \textbf{319} 1808

\bibitem{key-15} Rosenband T, Schmidt P O, Hume D B, Itano W M, Fortier
T M, Stalnaker J E, Kim K, Diddams S A, Koelemeij J C J, Bergquist
J C \textit{et al}. 2007 \textit{Phys. Rev. Lett.} \textbf{98} 220801 

\bibitem{key-16} Oskay W H, Diddams S A, Donley E A, Fortier T M,
Heavner T P, Hollberg L, Itano W M, Jefferts S R, Delaney M J, Kim
K \textit{et al}. 2006 \textit{Phys. Rev. Lett.} \textbf{97} 020801

\bibitem{key-17} Schmidt P O, Rosenband T, Langer C, Itano W M, Bergquist
J C and Wineland D J 2005 \textit{Science} \textbf{309} 749 

\bibitem{key-18} Giovannetti V, Lloyd S and Maccone L 2004 \textit{Science}
\textbf{306} 1330

\bibitem{key-19} Giovannetti V, Lloyd S and Maccone L 2002 \textit{Phys.
Rev. A} \textbf{65}, 022309

\bibitem{key-20} Taylor J M, Cappellaro P, Childress L, Jaing L,
Budker D, Hemmer P R, Yacoby A, Walsworth R and Lukin M D 2008 \textit{Nat.
Phys.} \textbf{4} 810

\bibitem{key-21} Maze J R, Stanwix P L, Hodges J S, Hong S, Taylor
J M, Cappellaro P, Jaing L, Gurudev Dutt M V, Togan E, Zibrov A S
\textit{et al}. 2008 \textit{Nature} \textbf{455} 644

\bibitem{key-22} Jiang L, Hodges J S, Maze J R, Maurer P, Taylor
J M, Cory D G, Hemmer P R, Walsworth R L, Yacoby A, Zibrov A S \textit{et
al.} 2009 \textit{Science} \textbf{326} 267

\bibitem{key-23} Geremia J M, Stockton J K, Doherty J C and Mabuchi
H 2003 \textit{Phys. Rev. Lett.} \textbf{91} 250801 

\bibitem{key-24} Stockton J K, Geremia J M, Doherty J C and Mabuchi
H 2004 \textit{Phys. Rev. A} \textbf{69} 032109 

\bibitem[27]{key-26} Bollinger J J, Itano W M, Wineland D J and Heinzen
D J 1996 \textit{Phys. Rev. A} \textbf{54} R4649 

\bibitem[28]{key-25} Gel'fand I M 1961 \textit{Lectures on Linear
Algebra (Interscience Tracts in Pure and Applied Mathematics)} ed
L Bers, R Courant \textit{et al.} (New York: Interscience Publishing
Inc.)

\bibitem[29]{key-27}Boixo S, Flamia S T, Caves C M and Geremia J
M 2007 \textit{Phys. Rev. Lett.} \textbf{98} 090401

\end{thebibliography}
\end{document}